\documentclass[twocolumn,showpacs,superscriptaddress]{revtex4}
\usepackage{dcolumn}
\usepackage{amsmath}

\def\he{$^{6}$He }

\makeatletter

\providecommand{\LyX}{L\kern-.1667em\lower.25em\hbox{Y}\kern-.125emX\@}

\usepackage[T1]{fontenc}
\usepackage[latin1]{inputenc}
\usepackage{graphics}
\usepackage{comment}
\makeatother

\begin{document}

\title{Improved di-neutron cluster model for $^6$He scattering}

\author{A.~M.\ Moro}
%\email{ariasc@us.es}
\affiliation{Departamento de F\'{\i}sica At\'omica, Molecular y
Nuclear, Facultad de F\'{\i}sica, Universidad de Sevilla,
Apartado~1065, E-41080 Sevilla, Spain}

\author{K. Rusek}
\affiliation{Department of Nuclear Reactions, 
Andrzej Soltan Institute for Nuclear Studies, Hoza 69, PL-00681 Warsaw, 
Poland}

\author{J.~M. Arias}
\affiliation{Departamento de F\'{\i}sica At\'omica, Molecular y
Nuclear, Facultad de F\'{\i}sica, Universidad de Sevilla,
Apartado~1065, E-41080 Sevilla, Spain}

\author{J.\ G\'omez-Camacho} 
%\email{gomez@us.es}
\affiliation{Departamento de F\'{\i}sica At\'omica, Molecular y
Nuclear, Facultad de F\'{\i}sica, Universidad de Sevilla,
Apartado~1065, E-41080 Sevilla, Spain}

\author{M. Rodr\'{\i}guez-Gallardo} 
%\email{gomez@us.es}
\affiliation{Centro de F\'{\i}sica Nuclear da Universidade de Lisboa,
 Lisbon, P-1649-003, Portugal}

\date{\today}

\begin{abstract}
The structure of the three-body Borromean nucleus $^6$He is approximated  by  a 
two-body di-neutron cluster model. The binding energy of the $2n$-$\alpha$ system is determined
to obtain a correct description of the $2n$-$\alpha$ coordinate, as given by a realistic three-body
model calculation.  
The model is applied to describe the break-up effects in 
elastic scattering of $^{6}$He on several targets, for which experimental data exist.
We show that an adequate description of the di-neutron-core 
degree of freedom  permits a fairly accurate description 
of the elastic scattering  of $^{6}$He on different targets. 
\end{abstract}

\pacs{24.10.-i, 24.10.Eq., 25.10.+s, 25.45.De, 25.60.Gc}

\maketitle

\vspace{2cm}
\section{Introduction}
The scattering of a weakly bound projectile by a target
represents a challenging as well as an interesting problem in nuclear 
physics. A proper understanding of the process requires an
accurate description of the 
%three-body 
structure of the projectile, including
all bound and unbound states that can be effectively coupled during 
the collision. 
%[REMOVE?:In practice, this amounts to solve a complicated quantum
%mechanical problem]  
In the case of weakly bound  two-body projectiles
the problem has been solved using the
Continuum Discretized Coupled--Channels (CDCC) method
\cite{Yah86,Kam86,Aus87}. Within the CDCC method, the reaction
process of a loosely bound two-body projectile by a structureless 
target is treated  within a three-body picture. The idea of the 
method is to represent the continuum part of the two-body projectile 
spectrum by a finite set of square integrable  states. These states
are then used to generate the diagonal as well as 
non-diagonal coupling potentials that enter the system of coupled equations. 
% To this end, 
% the continuum is divided  into a finite set of energy 
% intervals. For each interval, or bin, 
% a representative function is constructed by superposition 
% of true scattering wave functions
% within that interval. By construction, the set of functions 
% obtained in this way are 
% normalizable and mutually orthogonal. By projecting the  
% Schr\"odinger equation onto the bound and bin wave functions, 
% a set of coupled equations is
% obtained.  The method has been extremely successful and is 
% a reliable reference
% for any other alternative method.

In principle, the method can be extended to three-body projectiles. This will be 
the case, for instance, of Borromean nuclei, consisting of  three-body loosely bound and
spatially extended systems, typically composed of a compact core plus two
weakly bound neutrons ($n+n+c$), and with no bound binary subsystems. 
In this case, a description of the three-body spectrum of the 
projectile is required.  However, 
the calculation of the unbound spectrum of a three-body system is a very 
complicated problem by itself. In
general, each physical state will be a complicated superposition of many 
channels with all possible spin and orbital angular momenta configurations. 
% When used
% in a coupled-channels calculation, the fragment-target interactions have to be folded
% with the [COMPLETAR]. 
The calculation of the coupling potentials and the solution of the set of coupled equations 
in this large basis represents a complicated task. Despite these difficulties,
in two recent works \cite{Mat04a,Mat06a}, this method has been applied to describe the 
scattering of $^6$He on $^{12}$C and $^{209}$Bi. These calculations reproduce successfully 
existing elastic scattering data for these reactions and represent an important
advance towards the understanding of few-body nuclear reactions.

Most of the complexity of these processes involving  Borromean nuclei 
%(here generally denoted as $n+n+c$)
arises 
from the fact that these systems exhibit many excitation modes, which can be  associated with 
two different degrees of freedom: the  $n-n$ relative motion,  and the $(nn)-c$ motion. In general,
both modes will be excited during the collision. However, 
% In the sccattering of a three-body system, such as a Borromean nucleus ($n+n+c$), there 
% are two different modes that can be excited during the collision. One is 
% the $n-n$ relative
% motion and, the other, the $(nn)$-$c$ motion.  
% In a scattering process, however, not all degrees of freedom of the weakly bound nucleus
% will be equally relevant. In particular, 
when the system is scattered by a medium mass or heavy target,
the projectile-target interaction will excite mainly the coordinate between
the neutrons and the core, since the repulsive Coulomb interaction will tend to repel 
the charged core, while the neutrons can approach closer to the target. Moreover, the nuclear 
interaction will attract more strongly the weakly bound neutrons. So, the net effect of the 
interaction with the target will be to stretch the $nn-c$ coordinate, pushing the core 
apart from the target and pulling the neutrons close to it. Thus, a description of the 
projectile excitation mechanism that takes into account explicitly the $nn-c$ coordinate should 
explain the main features of the reaction mechanism of the three-body system with the target. 
% The projectile will be 
% then stretched in the $(nn)$-$c$ coordinate.  Within this
% picture, it is expected that an accurate description of the 
% relative motion between 
% the neutrons and the core will suffice to describe, at least approximately, the
% main features of the  scattering process. 

Given the complexity of the full CDCC  calculations with three-body projectiles, 
the development of these 
simple models can be very helpful to understand the main features of these processes 
by retaining only the essential ingredients to keep the model realistic.

% In this work we revisit the so called di-neutron model for the $^6$He  case. We review its 
% main features and limitations, and we propose a method to improve the accuracy of the 
% model, while keeping its simplicity. The new method is tested against 
% existing experimental data for $^6$He scattering on several targets. 

In this work we revisit the so called di-neutron model for the $^6$He  case. In Sec.~\ref{sec:3b}, 
we address the problem of the $^6$He structure within a three-body model. In 
Sec.~\ref{sec:2b}, we review the di-neutron model for 
this nucleus,  and we propose a method to improve the accuracy of the 
model, while keeping its simplicity. In Sec.~\ref{sec:calculos}
the new method is tested against  existing experimental data for $^6$He scattering on several targets. Finally, 
Sec.~V is for summary and conclusions.

\section{Three-body model for $^6$He \label{sec:3b}}
Within a three-body picture, the wavefunction of the  $^6$He system can be 
conveniently expressed in terms of one of the Jacobi sets of coordinates. For the purposes of 
the present work, the most suitable representation is that in terms of the neutron-neutron 
relative coordinate, {\bf x}, and the $nn$-$^4$He coordinate, {\bf y}. This wavefunction, 
here denoted  $\Psi^{3B}(\mathbf{x},\mathbf{y})$,
can be obtained by solving the Schr\"odinger equation, using any of the methods  
proposed in the literature. Here, we followed 
the procedure proposed in \cite{Zhu93,face}, in which the wavefunction is expanded 
in hyperspherical coordinates. The basic ingredient of the calculation are the two-body 
interactions between the subsystems ($n-n$ and $n-\alpha$). 
Besides the two-body 
potentials, the model Hamiltonian also includes a simple central
three-body force depending on the hyperradius. This is introduced to overcome the
under-binding caused by the other closed channels, such as the 
$t$+$t$ channel. The $n$-$^4$He potential
is taken from Refs.~\cite{Bang79,Tho00}, with central and
spin-orbit components, and the neutron-neutron potential, 
with central, spin-orbit and tensor components, from the prescription of
Gogny, Pires and  Tourreil \cite{gpt}. These
calculations were performed with the code STURMXX 
\cite{sturmxx} 
which uses the formalism described in 
\cite{Tho00}. The maximum
hyperangular momentum was set to $K_{\max}=20$
and the three-body force was adjusted to give the right binding
energy. The calculated three-body wave function has a binding
energy of $\epsilon_b=0.955$~MeV and a rms point nucleon matter radius of
$2.557$~fm when assuming an alpha-particle rms matter radius of
$1.47$~fm. Further details of these 
calculations can be found in Ref.~\cite{manoli05,Tho00}.
It should be noted that the three-body wavefunction    is a complicated 
superposition of many channel 
configurations. Each channel is characterized by the angular momentum 
in the $n$-$n$ and $(nn)$-$\alpha$ coordinates ($l_x$ and $l_y$), the total 
orbital angular momentum ($L$) and  the total 
spin of the neutron pair ($S_x$).  
% It should be noticed that the 
% three-body WF $\Psi(\mathbf{x},\mathbf{y})$ is a complicated superposition of many 
% channel configurations, involving the quantum numbers $l_x$, $l_y$, $S_x$ and  $J$. 
In the
$^6$He ground state the dominant configuration corresponds to $S_x=l_x=l_y$=0, which 
contributes to 80\% of the norm. 
%For simplicity, we will consider only this configuration 
%and, accordingly, the di-neutron model will have also $l_y=S_x$=0. 

% --------------------------------------------------------------
% NEUTRONS DENSITY 
% --------------------------------------------------------------
\begin{figure}[t]
{\par\centering \resizebox*{0.45\textwidth}{!}
{\includegraphics{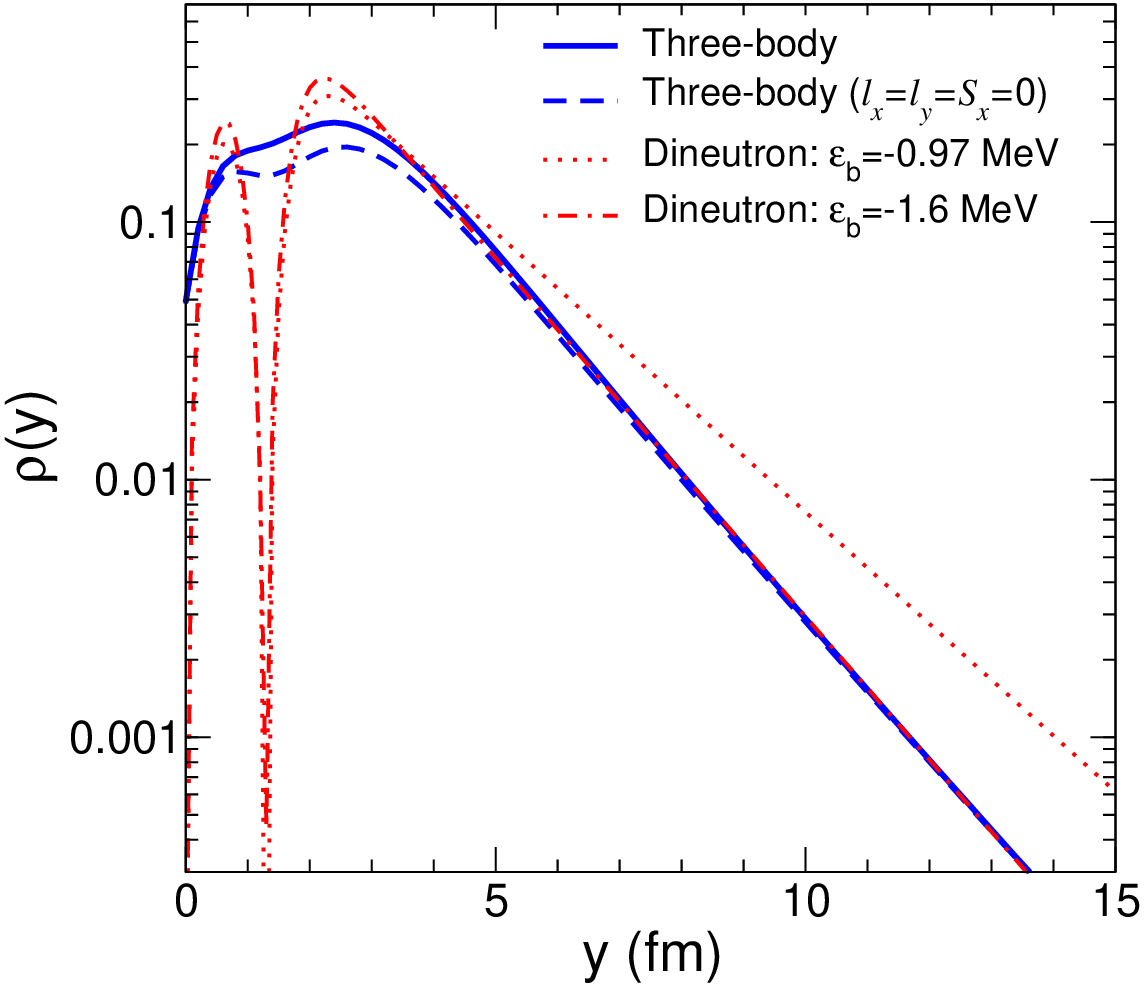}}\par}
\caption{\label{fig:rhoy} (Color online) 
Neutron density in the di-neutron model,  compared with a realistic 
three-body model. The di-neutron calculations use a Woods-Saxon potential with 
radius $R=1.9$~fm and diffuseness $a=0.25$~fm. }
\end{figure}
% --------------------------------------------------------------

In order to compare with the  di-neutron model, presented below, we
%For the purpose of comparing with the di-neutron model, presented below, we 
consider now the behavior of the 
wavefunction in the ${\bf y}$ coordinate. For this purpose, we calculate
the probability density in the 
$(nn)$-$\alpha$ relative coordinate, $y$, here denoted as $\rho(y)$. This was calculated by 
integrating the square of the three-body wavefunction on the neutron-neutron coordinate, i.e.\
\begin{equation}
\label{eq_rho3b}
\rho^{3B}(y)= y^2 \int |\Psi^{3B}(\mathbf{x},\mathbf{y})|^2 
d\mathbf{x} d\Omega_y \,\, ,
\end{equation}
where $\Psi(\mathbf{x},\mathbf{y})$ is the total three-body wavefunction and 
$\Omega_y$ denotes the angular variables $(\theta_y, \phi_y)$. This density is 
plotted in  Fig.~\ref{fig:rhoy} with the thick solid line. To illustrate the dominance of the 
 $S_x=l_x=l_y=0$ component, we show also the same quantity, retaining only this component in
the wavefunction (thick dashed line). It can be noticed that, for $y> 5$ fm, the di-neutron density 
is completely determined by this component. Consequently, a realistic model for the \he ground 
state wavefunction must account, at least, for this configuration.

% ---------------------------------------------------------------
% B(E1) B(E2)
% ------------------------------------------------------------------------
\begin{figure}[t]
{\par\centering \resizebox*{0.45\textwidth}{!}
{\includegraphics{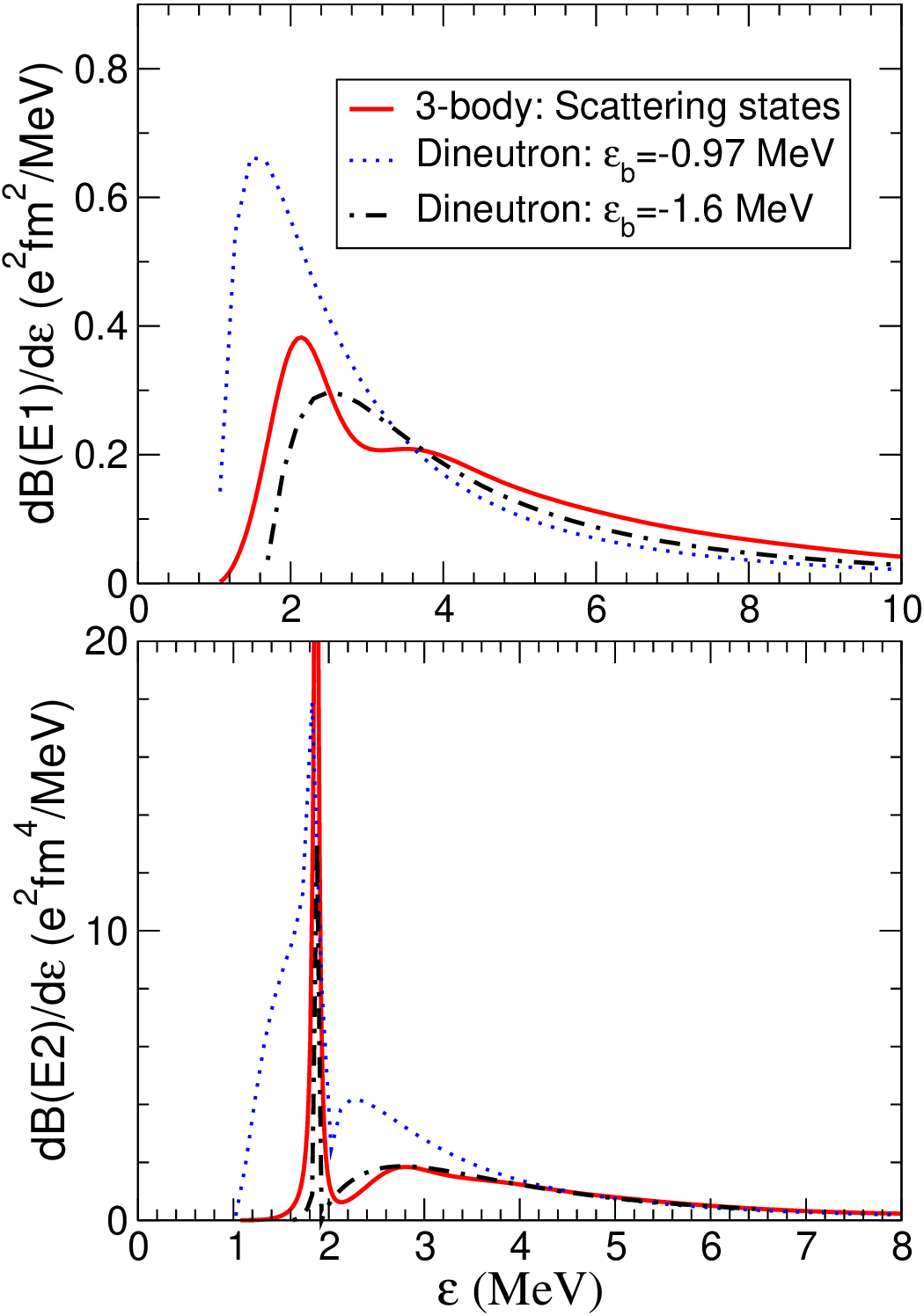}}\par}
\caption{\label{fig:be12} (Color online) 
$B(E1)$ and $B(E2)$ transition strengths for the $^{6}$He nucleus using different 
structure models. The solid lines correspond to the three-body calculation 
obtained with the {\it true} scattering states \cite{Tho00}. Di-neutron calculations 
using for the $2n$-$\alpha$ binding energy either the two-neutron separation energy (dotted-line)
or the modified value $\epsilon_b=-1.6$~MeV (dotted-dashed line) are also shown.
}
\end{figure}
% -------------------------------------------------------------------------

In scattering calculations involving the  \he nucleus it is essential 
to include also a realistic description of the continuum 
states, given the large breakup probability of weakly bound nuclei. In the case of the 
Coulomb interaction, 
the response of the continuum to excitations of multipolarity $\lambda$ is conveniently treated
in terms of the reduced transition probability, $B(E\lambda)$ \cite{Bri94}. In Fig.~\ref{fig:be12}
we consider the  $B(E1)$ (upper panel) and $B(E2)$ (bottom panel) 
distributions, plotted as 
a function of the excitation energy of the \he nucleus  with respect to the ground state. In 
both panels the full three-body calculation  is 
depicted by the thick solid line. In these calculations, the continuum states were 
represented by true scattering 
wave functions, as reported in Ref.~\cite{Tho00}. The narrow peak in the $B(E2)$ corresponds
to the known $2^+$ low lying resonance.

\section{The di-neutron cluster model of $^6$He \label{sec:2b}}

We want to consider situations in which the $(nn)-c$ degree of freedom is more relevant than the $nn$ degree of freedom in
$^6$He. This is the case, for example, when electric operators are considered in structure calculations, or when 
Coulomb forces dominate in a collision. One could think then in approximating the three-body
wavefunction by a product of two-body wavefunctions, i.e. 
%$\Psi^{3B}(\mathbf{x},\mathbf{y})$ 
\begin{equation}
\label{eq_3b2b}
\Psi^{3B}(\mathbf{x},\mathbf{y}) \simeq \Psi^{2B} (\mathbf{y}) \psi(\mathbf{x}) .
\end{equation}

The {\it di-neutron} model  takes into account only the $(nn)-c$ degree of freedom, whereas 
 the relative motion between the 
two valence neutrons is ignored. This amounts to consider that the neutron pair remains in 
a highly correlated state $\psi(\mathbf{x})$ during the collision and, hence, excitations  
in this coordinate are not permitted. Moreover, the neutrons are assumed to be coupled to 
spin zero, and bound to  an inert $\alpha$ core  in a $s$-wave relative motion which, 
as we have seen before, is the 
dominant configuration in the $^{6}$He ground state wavefunction.
This model is inspired in the deuteron-$\alpha$ cluster model, which has been very 
successful in the description of $^{6}$Li scattering within the continuum discretized 
coupled channels method (eg.~\cite{Sak82,Sak87,Kee96}).
% However, previous attempts to describe $^6$He scattering by
% heavy targets \cite{Rus03a,kee03a} using this simple model showed that this simplified description of the $^6$He 
% nucleus tends to overestimate the effect of the 
% continuum couplings. In \cite{Rus05a,Mat06a} it was shown that this 
% failure is mainly due to an
% overestimation of the $E1$ strength and, indeed, by reducing the strength of the 
% dipole couplings the agreement with the data could be significantly improved \cite{Rus05a}.

The problem in the di-neutron model lies on evaluating the wavefunctions, for the bound and continuum states, 
 describing the motion of the halo neutrons relative to the 
$\alpha$ core, $\Psi^{2B}(\mathbf{y})$. Following the cluster model, one assumes that these 
wavefunctions can be obtained
as the eigenstates of the Hamiltonian  corresponding to a certain  $2n$-$\alpha$ interaction. 
Typically, one  assumes some reasonable geometry for the $2n$-$\alpha$ interaction, and then adjusts the potential depth
to obtain a given binding energy for the $2n$-$\alpha$ system. 

In all the calculations
here presented, the $\alpha+2n$ interaction was parameterized using a standard Woods-Saxon 
form, with radius $R_0$=1.90~fm and diffuseness $a=0.25$~fm, which corresponds to the 
set III of Ref.~\cite{Rus01}. The ground state wavefunction was assumed to 
be a pure 2S configuration,
%the $2n$-$\alpha$ motion was assumed to be on a pure 2S state  
since, due to the Pauli principle, the 1S state is forbidden.

So, the key question is:  which is the binding energy that
one should use  for the $2n$-$\alpha$ system, so that the corresponding wavefunction gives a reasonable description of 
$^6$He in a di-neutron model?

In the application of the  deuteron-$\alpha$ cluster model to $^{6}$Li \cite{Sak82,Sak87,Kee96}, one evaluates the binding energy
for $d-\alpha$ just as the separation energy of $^{6}$Li into $d+\alpha$. 
This is a reasonable procedure, which can be applied because the deuteron is bound
by 2.2 MeV, which is more than the separation energy of  $^{6}$Li into $d+\alpha$, and so one can argue that the relative wavefunction of
the valence proton and neutron within   $^{6}$Li is not very different from that in a free deuteron. Besides, the  deuteron-$\alpha$ 
cluster model gives  reasonable values for the mean square radii of  $^{6}$Li.

In the applications of the di-neutron model done so far to $^6$He  \cite{Rus01,Mac03,Rus04,Mac04,Gio05,Rus05a}, 
the binding energy of the di-neutron has been taken as the two-neutron separation energy of $^6$He, i.e., $|\epsilon_b|$=$S_{2n}$=0.975~MeV. 
It should be noticed that, with this choice, one is assuming implicitly that the relative wavefunction of two neutrons within
$^6$He would be in a state similar to that of two neutrons with zero relative energy. This leads  
to an unrealistic wavefunction for the di-neutron-$\alpha$ motion, as discussed below. 
% ----------------------------------------------------------------------------------------

The use of a binding energy  $\epsilon_b=-0.975$~MeV
yields the potential depth $V_0=93.51$~MeV. 
For the $\ell=0$ and $\ell=1$ continuum states we used
the same potential as for the ground state. 
For the $\ell=2$ continuum, the potential
depth was  changed to $V_0=91.25$ MeV, in order to get the $2^+$ resonance at the 
% correct excitation energy, with respect to the ground state. 
value obtained in the
the three-body calculation which, in turn, is close to the experimental value 
($\epsilon_x=0.825$ MeV above the breakup threshold).

To illustrate this, we  compare the density probability associated with 
the  $nn-c$ coordinate in the two- and three-body models.  In the di-neutron model, 
the neutron density analogous to 
Eq.~(\ref{eq_rho3b}) is  simply obtained as $\rho^{2B}(y)= |y R(y)|^2$, where 
$R(y$) is the radial part of the wave function $\Psi^{2B}$. In Fig.~\ref{fig:rhoy}, 
the density probability obtained with this  model
is given by the dotted line. When compared to the realistic  three-body calculation (thick solid line) 
it becomes apparent that the former extends to considerably larger distances. For example, the 
rms associated to the di-neutron-$\alpha$ coordinate is 4.36~fm, considerably larger than the prediction 
of the three-body model, 3.25~fm. In view of this result it is 
not surprising that the coupling of the ground state wavefunction with the dipole continuum states
is unphysically enhanced in the two-body model. 

This is indeed the case, as we can see in Fig.~\ref{fig:be12}.  In both panels,
the dotted line corresponds to the di-neutron model. These distributions clearly overestimate 
both the  $E1$  and $E2$ strengths predicted by the three-body model (thick solid lines). Not 
surprisingly, previous attempts to describe $^6$He scattering by
heavy targets \cite{Rus03a,kee03a} using this model showed that this simplified description of the $^6$He 
nucleus tends to overestimate the effect of the 
continuum couplings. In \cite{Rus05a} it was shown that, 
% this  failure is mainly due to an
% overestimation of the $E1$ strength and, indeed, by 
reducing the strength of the 
dipole couplings, the agreement with the data could be significantly improved.

%%%%%%%%%%%%%%%%%%%%%%%%%%%%%%%%%%%%%%%%%%
%\section{An improved di-neutron model of $^6$He \label{sec:2bb}}
%%%%%%%%%%%%%%%%%%%%%%%%%%%%%%%%%%%%%%%%%%%%%

%From the considerations above, it becomes clear that the common procedure to generate the 
%$2n$-$\alpha$ wavefunction in the di-neutron model leads to an unrealistic model.  

We consider that the $2n$-$\alpha$ binding energy  used in the di-neutron model should not be given by the two-neutron
separation energy. The di-neutron system which appears in $^6$He is a correlated state, which, in the absence of the
$\alpha$ particle, will be given by a wave packet with  positive expectation value of the energy. 
Thus,  the actual $2n$-$\alpha $ binding energy should be more negative, to compensate for the positive energy of the di-neutron. 
We propose to obtain   $2n$-$\alpha$ binding energy  to  reproduce the known properties of the $^6$He system, such as the rms radius and the transition strengths, within the di-neutron model. 
Asymptotically, the di-neutron wavefunction behaves  as $\propto\exp(-k y)$ with 
$k=\sqrt{2 \mu |\epsilon_b|}/\hbar$ and $\mu$ the  $2n$-$\alpha$ reduced mass. Then, a  
natural choice for $|\epsilon_b|$ is to make the slope as close as possible to the three-body 
case. This leads to the value $|\epsilon_b| =1.6$~MeV. The density calculated with 
this value, shown by   the dotted-dashed line in Fig.~\ref{fig:rhoy}, reproduces very 
well the three-body calculation (thick solid line) for separations beyond 4 fm. The 
di-neutron-$\alpha$  mean square separation obtained with the new wavefunction is reduced to 
% the way to correct the asymptotic behavior of 
% the di-neutron model is to increase the binding energy in order to make the slope as similar as 
% possible to that of the 
% three-body case. We found that, using the two-neutron separation energy 
% $S_{2n}= |\epsilon_b| =1.6$~MeV, 
% the tail of the three-body wavefunction in the $y$ coordinate is very well reproduced for 
% distances beyond 4 fm. This is shown  by the dotted-dashed line in Fig.~\ref{fig:rhoy}. With the 
% new value of $S_{2n}$, the $2n$-$\alpha$ rms is reduced to 
3.4~fm, in much better agreement with 
the three-body result. With the new binding energy, the depth of the $s$-wave is modified 
to $V_0$=96.06~MeV. Again, this depth was used for the $p$-waves. The depth of the $\ell=2$ potential
had to be changed to  $V_0$=92.7~MeV, in order to get the $2^+$ resonance at the correct excitation 
energy with respect to the ground state. 

We note that, among the three geometries proposed in  Ref.~\cite{Rus01}, namely, the set I ($a=0.65$~fm), 
set II ($a=0.39$~fm) and III ($a=0.25$~fm) the latter is found to reproduce more accurately the three-body 
density. The other two geometries, having a larger diffuseness, give rise to a higher rms, even after 
modification of the two-neutron separation energy to correct the slope of the di-neutron density. 

It can be seen that, with this geometry and binding energy, the $E1$ and $E2$ transition strengths are 
also well reproduced. This is shown in 
Fig.~\ref{fig:be12} by the dotted-dashed lines. 
%, we study the effect of the modification on the separation energy on the 
% transition probabilities. The 
% dotted-dashed line is the di-neutron model
% calculation, using the modified value $S_{2n}=1.6$~MeV. 
It 
is observed that this increase of the binding energy reduces both strengths
showing a much better agreement with the prediction of the three-body 
calculation. Note that, due to the modification of the binding energy, 
the breakup threshold appears at a higher energy in our di-neutron model. So, the di-neutron model does not describe the low-energy
continuum of $^6$He, which is below 1.6 MeV excitation energy. However, it describes fairly well the continuum around the maximum
of the $B(E1)$ distribution (2 MeV) and beyond.

%In the next section, we test this model in scattering calculations. 

% ----------------------------------------------------------------------------
% ELASTIC SCATTERING CALCULATIONS
% -------------------------------------------------------------------------------
\section{Calculations \label{sec:calculos}}

In the remaining, we compare  the two-body models discussed above 
with the elastic scattering 
data for several reactions induced by $^6$He.  In all the calculations here presented, we used the 
geometry of the $2n$-$\alpha$ potential with the smaller diffuseness ($a=0.25$~fm). All these calculations are 
performed within the standard CDCC method \cite{Yah86}.

% --------------------------------------------------------------------------
% 6He+64Zn elastic scattering
% --------------------------------------------------------------------------
\begin{figure}[t]
{\par\centering \resizebox*{0.45\textwidth}{!}
{\includegraphics{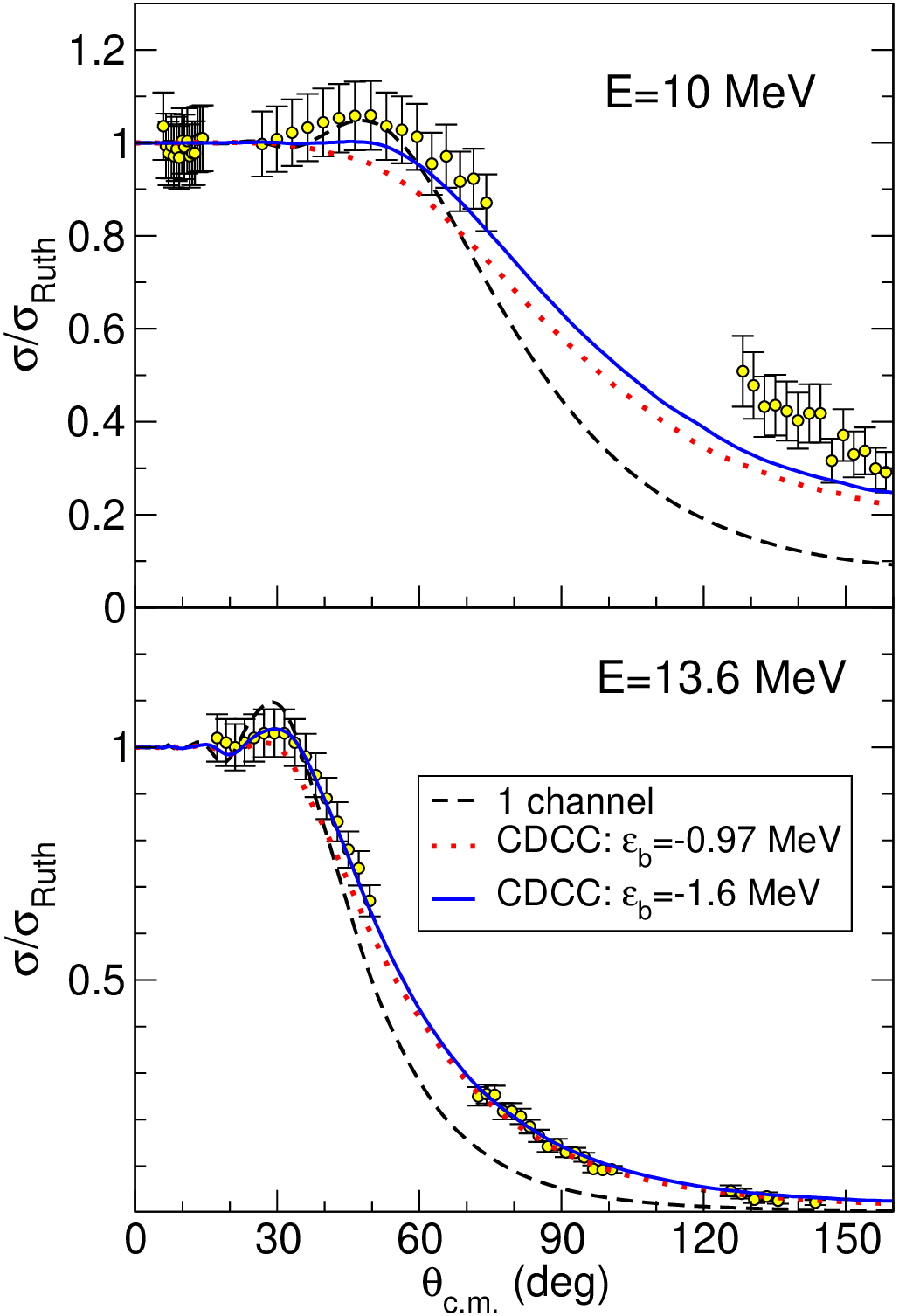}}\par}
\caption{\label{fig:he6zn} (Color online) 
Elastic scattering angular distribution, divided by the Rutherford cross section, 
for the reaction  $^{6}$He+$^{64}$Zn at $E=10$ and 13.6 MeV. The dashed line is 
the cluster-folding calculation without inclusion of the continuum. The dashed
line and the thick solid line are the di-neutron model calculations using the
binding energy  $\epsilon_b=S_{2n}=-0.975$~MeV, and  the modified binding energy $\epsilon_b=-1.6$ MeV, respectively.
The experimental data are taken from \cite{DiPiet04}. 
}
\end{figure}
% ----------------------------------------------------------------------------

We first consider the reaction $^{6}$He+$^{64}$Zn at Coulomb barrier energies, 
which was recently measured by Di Pietro {\it et al.} \cite{DiPiet04}.  The 
$^{6}$He (=$2n+\alpha$) continuum was discretized into $N=7$ energy bins, evenly
spaced in the asymptotic momentum $k$, and  up to a maximum
excitation energy of $\epsilon_\mathrm{max} = 7$ MeV. We included $s$, $p$ and $d$ waves for 
the  $2n$-$\alpha$ relative orbital angular momentum. Inclusion of $f$ waves had a negligible effect on
the elastic angular distributions. 

In these calculations, the $\alpha$+$^{64}$Zn interaction was taken from the optical model fit 
performed in  \cite{DiPiet04}. For the $2n + ^{64}$Zn interaction, we used the 
parameters of  the d+$^{56}$Fe potential obtained in Ref.~\cite{AlQ00}. Diagonal as well
as non-diagonal potentials were derived from these potentials by means of a single-folding 
method, as described elsewhere \cite{Rus01}. The coupled equations were integrated up to 
100 fm, and using 50 partial waves for the projectile-target relative motion. These calculations
were performed with the code FRESCO \cite{fresco}.

In Fig.~\ref{fig:he6zn} we show the results for the elastic scattering angular
distribution at 
the laboratory energies  $E$=10 and 13.6 MeV, along with the data of Di Pietro
{\it et  al.} \cite{DiPiet04}. For each energy three curves are shown: the dashed 
line is the cluster--folded calculation in which the projectile-target interaction 
is folded with the ground-state density of the $^{6}$He nucleus, without 
inclusion of the continuum. At the higher energy this calculation exhibits a 
pronounced rainbow
which is not observed in the data. Moreover, at both energies these calculations clearly
underestimate the data at backward angles. Inclusion of the continuum within the conventional
di-neutron model (dotted lines) improves the agreement at backward angles, but
reduces too much the cross section at the rainbow, thus underpredicting the data. 
This effect is a direct consequence of the overprediction of the $B(E1)$ distribution 
in the di-neutron model, as explained above.  Finally,
the thick solid line is the CDCC calculation with the di-neutron model with a modified
binding energy ($\epsilon_b=-1.6$ MeV). This calculation improves the agreement at the rainbow, 
particularly at $E=13.6$ MeV. At  $E=10$~MeV this calculation slightly  
underestimates the data at backward angles, but we could not find an explanation for this discrepancy.

\begin{figure}[t]
{\par\centering \resizebox*{0.45\textwidth}{!}
{\includegraphics{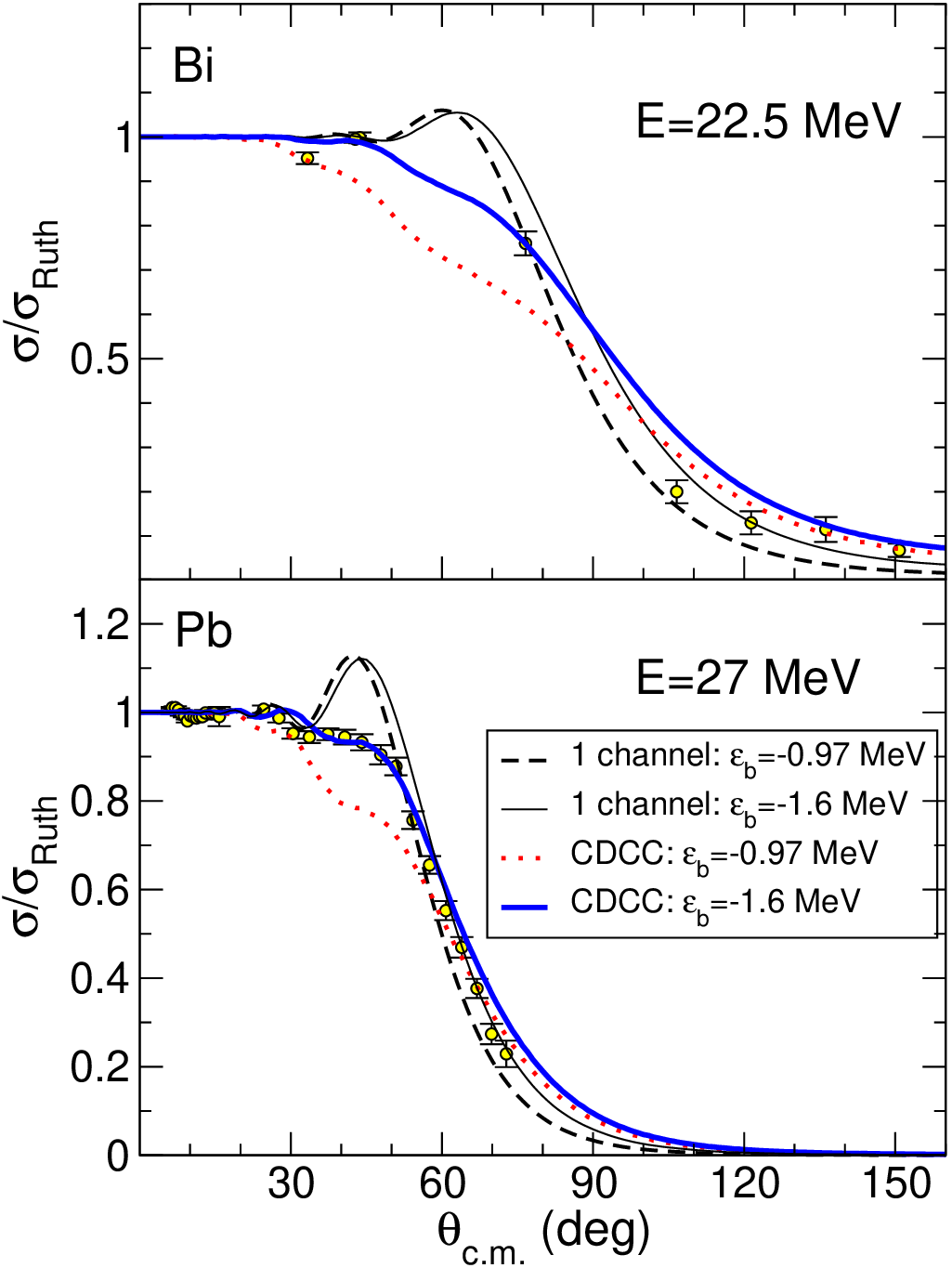}}\par}
\caption{\label{fig:he6bi} 
Elastic scattering angular distribution, divided by Rutherford cross section,
for the reactions  $^{6}$He+$^{209}$Bi at $E=22.5$~MeV (upper part) 
and $^{6}$He+$^{208}$Pb at $E=27$~MeV (lower part). The dashed line is
the calculation without inclusion of the continuum, using the binding 
energy $\epsilon_{b}=-0.97$ MeV. The thin solid line is a similar calculation, 
but with $\epsilon_{b}=-1.6$ MeV. The dotted
line and the thick solid line are the di-neutron model calculations using 
the binding energy  $\epsilon_b=S_{2n}=-0.975$~MeV, and  the
modified binding energy $\epsilon_{b}=-1.6$ MeV, respectively.
Experimental data are from \cite{Kak03} and \cite{Agu00}.
}
\end{figure}

% --------------------------------------------------------------------------
% 6He + Pb,Bi elastic scattering
% --------------------------------------------------------------------------
Next, we study the  $^{6}$He+$^{208}$Pb,$^{209}$Bi,$^{197}$Au reactions, which were
recently analyzed in \cite{Rus05a}. In Fig.~\ref{fig:he6bi} 
we present the calculations for the Pb and Bi targets, along with the experimental 
data from Kakuee {\emph et al.} \cite{Kak03} and Aguilera {\emph et al.} \cite{Agu00}, 
respectively. Details of the fragment-target optical potentials and  binning scheme can 
be found in  Ref.~\cite{Rus05a}.
The meaning of the lines is the same 
as in Fig.~\ref{fig:he6zn}.  In both cases, the calculation without continuum displays a marked rainbow, 
which is not observed in the data. We have added also a calculation without continuum, but 
with the modified binding energy  $\epsilon_{b}=-1.6$ MeV (thin solid line). This calculation illustrates the {\em static} effect caused by the change in the ground state wavefunction produced by the 
modification of the binding energy. Qualitatively, these two calculations are very similar. In particular, the pronounced rainbow is still present in the new calculation. The similitude between these two 
calculations indicates that the trend of the data can not be simply explained by changing the size of 
the di-neutron$-\alpha$ wavefunction and, consequently, dynamical effects are indeed very important. 
Inclusion of the continuum within the conventional di-neutron model (dotted
 line), produces a strong reduction of the cross section at intermediate angles, largely underestimating
the data.  As 
noted above, this is caused by the  
 overestimation of the dipole couplings in this model. In  
the modified di-neutron model the rainbow is also suppressed, but the final result is in very good 
agreement with the data.

% ----------------------------------------------------------------------------

% --------------------------------------------------------------------------
% 6He+197Au elastic scattering
% --------------------------------------------------------------------------
Finally, we discuss the results for the $^{6}$He+$^{197}$Au reaction at $E=27$, 29 and 40 MeV, 
and compare with the data of \cite{Raa01}. The 
lines have the same meaning as in figures \ref{fig:he6zn} and \ref{fig:he6bi}. Similarly to 
the case of the lead target,  at $E=27$, 29 MeV the one channel calculation exhibits a pronounced
 rainbow, which is almost absent  in the data. This effect is very well accounted for in the modified 
di-neutron calculation. At $E=40$~MeV, the rainbow is suppressed somewhat, but not completely, in the full CDCC calculation. The lack of data at the relevant angles does not permit to 
make strong conclusions about the existence of the rainbow, but the agreement between the data and 
the calculation is fairly good where the comparison is possible. 

It is interesting to note that the underestimation of the data in the conventional di-neutron model 
is more pronounced at lower energies. This is because at lower energies dipole Coulomb couplings 
become more important and,  
as we showed before, these couplings are unphysically enhanced in the conventional di-neutron model.

\begin{figure}[t]
{\par\centering \resizebox*{0.45\textwidth}{!}
{\includegraphics{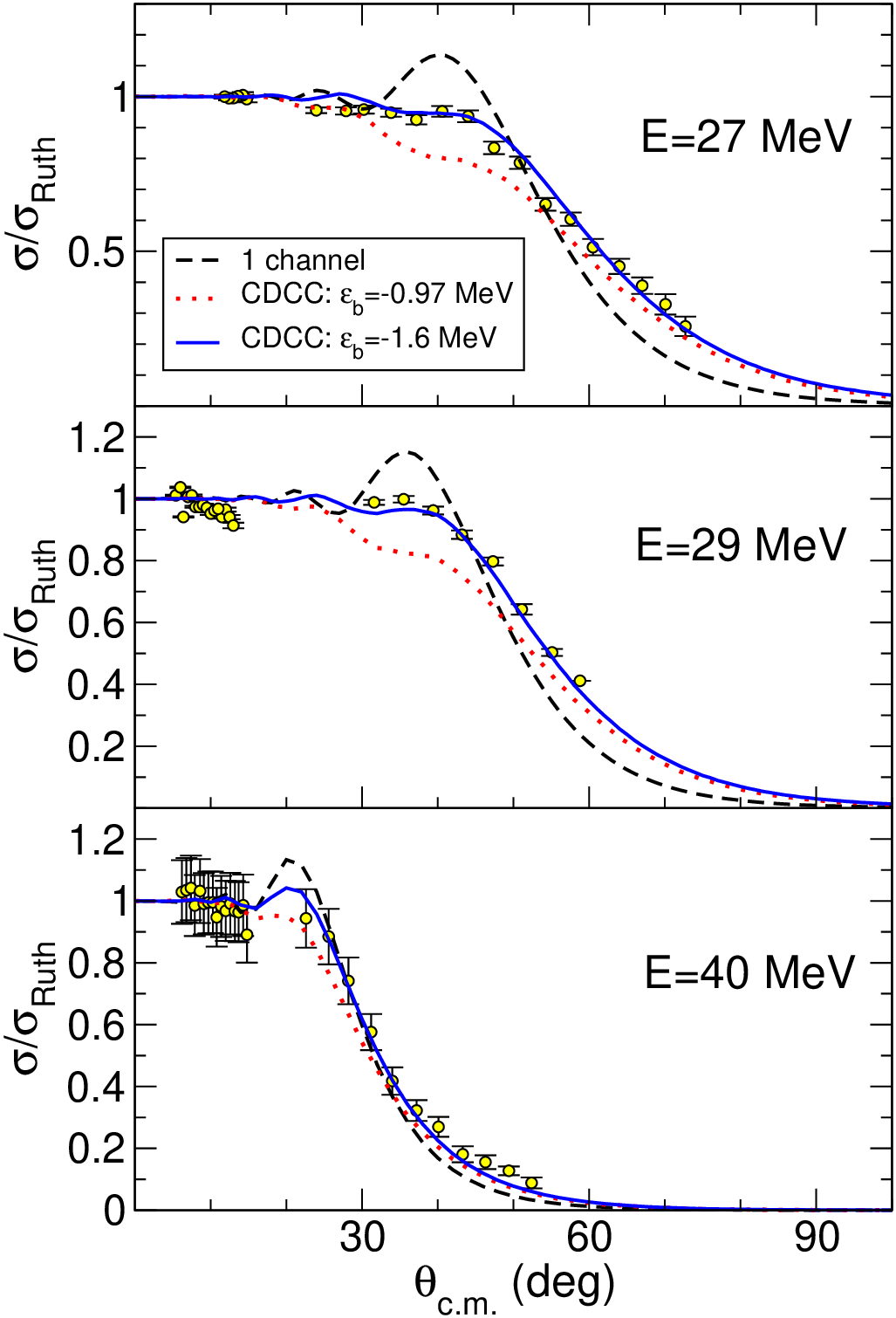}}\par}
\caption{\label{fig:he6au} 
Elastic scattering angular distribution, divided by the Rutherford cross section,
for the reaction  $^{6}$He+$^{197}$Au at $E=27$, 29 and 40~MeV. 
The dashed line is
the cluster-folding calculation without inclusion of the continuum. The dotted
line and the thick solid line are the  calculations using the di-neutron model with 
 $\epsilon_{b}=-S_{2n}=-0.975$~MeV  and 
% binding energy, and with the modified separation energy 
$\epsilon_{b}=-1.6$ MeV, respectively.
The experimental data are taken from \cite{Raa01}.
}
\end{figure}
% ----------------------------------------------------------------------------

% ------------------------------------------------------------------------
% DISCUSSION
% --------------------------------------------------------------------------
All these calculations show that the proposed model describes fairly well the elastic data for 
different targets and could even be used as a predictive tool for reactions for which data 
do not exist. The good agreement with the data clearly supports the idea, anticipated in 
the introduction, that the degree of freedom which enters actively in the elastic scattering 
of Borromean systems on medium mass and heavy targets, is that for the relative motion between 
the halo neutrons and the core.

\section{Summary and conclusions \label{sec:summary}}

In this work,  we have studied the application of  the di-neutron model to describe $^6$He structure and scattering.
We find that, when the di-neutron model is applied assuming for the $2n$-$\alpha$ binding energy the two-neutron
separation energy of $^6$He, the description of the structure of $^6$He obtained is not in agreement with the results of 
a realistic three-body calculation.  One obtains an  unrealistically long  tail of the $2n$-$\alpha$ relative 
wavefunction, and too large values of the $B(E1)$ and $B(E2)$ distributions. 
When this model of $^6$He is used in scattering calculations, the couplings between the ground state and 
the continuum states are overestimated, and this produces too much absorption from the elastic channel.

 We have proposed a modified di-neutron model, in which the 
% the di-neutron model, determining the 
$2n$-$\alpha$ binding energy is set to reproduce the 
density distribution in the $2n$-$\alpha$ coordinate given in a 
realistic 3-body model. We find that a $2n$-$\alpha$ binding energy 
of 1.6 MeV produces  a rms and  $B(E1)$ and $B(E2)$ distributions which are similar to  those obtained in a realistic three-body  calculation. 

The model has been tested for several reactions induced by $^6$He, providing 
in all cases a very good description 
of the elastic scattering data. These results indicate that, despite its simplicity, the model 
can provide a useful and reliable description of reactions involving the $^6$He 
nucleus. Using 
an identical procedure, the method could be also extended to other Borromean systems, such as 
$^{11}$Li or $^{14}$Be.

We would like to emphasize that the present model is not intended to replace the realistic
three-body calculations for the scattering of  Borromean nuclei. The development of these models, 
although numerically more demanding, are of great importance for a full quantitative 
understanding of these processes. However, we believe that simple models, as those discussed 
here, are also very useful  to provide  us with a transparent physical interpretation of 
these collisions.

\begin{acknowledgments}
This work has been partially supported by the Spanish Ministerio de
Educaci\'on y Ciencia and under the projects  FPA2005-04460 
and FIS2005-01105.  
M.R.G. acknowledges financial support by
FCT under the grants POCTI/ISFL/2/275 and
POCTI/FIS/43421/2001.
A.M.M. acknowledges financial support  from the Junta de  Andaluc\'{\i}a. We 
are grateful to A.\ Di Pietro and C.\ Angulo for useful information concerning  the
$^6$He+$^{64}$Zn data. 
\end{acknowledgments}

%Standard
%\bibliographystyle{unsrt}
%\bibliography{refer}

%APS  
\bibliographystyle{apsrev}
\bibliography{dineutron}
%\bibliography{tho}

\end{document}